\documentclass[11pt]{article}
\usepackage{moriond,epsfig}

\def\Journal#1#2#3#4{{#1} {\bf #2}, #3 (#4)}

\def\PLB{{\em Phys. Lett.}  B}
\def\PRL{\em Phys. Rev. Lett.}
\def\PRD{{\em Phys. Rev.} D}

\def\MPL{\em Mod.Phys.Lett.}
\def\CQG{\em Class. Quantum Grav.}
\def\IJMP{\em Int. Journ. Mod. Phys.}

\def\be{\begin{equation}}
\def\ee{\end{equation}}
\def\bea{\begin{eqnarray}}
\def\eea{\end{eqnarray}}

\begin{document}
\vspace*{4cm}
\title{Scale Invariance, Inflation and the Present Vacuum Energy of the Universe}

\author{E.I.Guendelman }

\address{Department of Physics, Ben Gurion University,\\
Beer Sheva 84105, Israel}

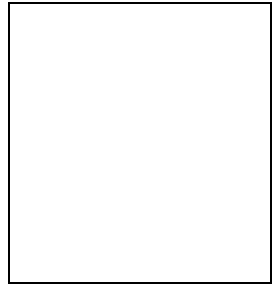
\maketitle\abstracts{                                                              
The possibility of mass in the context of scale-invariant, generally          
covariant theories, is discussed. Scale invariance is considered  in          
the context of a gravitational theory where                                   
the action, in the first order formalism, is of the form $S =                 
\int L_{1} \Phi d^4x$ + $\int L_{2}\sqrt{-g}d^4x$ where $\Phi$ is a           
density built out of degrees of freedom independent of the metric.            
For global scale invariance, a "dilaton"                                      
$\phi$ has to be introduced, with non-trivial potentials $V(\phi)$ =          
$f_{1}e^{\alpha\phi}$ in $L_1$ and $U(\phi)$ = $f_{2}e^{2\alpha\phi}$ in      
$L_2$. This leads to non-trivial mass generation and a potential for          
$\phi$ which is interesting for inflation. The model after ssb can be 
connected to the induced gravity model of Zee, which is a successful
model of inflation. Models of the present universe and a natural 
transition from inflation to a slowly accelerated universe at late
times are discussed.}

\section{The Model in the absence of fermions}

The concept of scale invariance appears as an attractive possibility for a    
fundamental symmetry of nature. In its most naive realizations, such a        
symmetry is not a viable symmetry, however, since nature seems to have        
chosen some typical scales.                                                   
                                                                              
Here we will find that scale invariance can nevertheless be incorporated      
into realistic, generally covariant field theories. However, scale            
invariance has to be discussed in a more general framework than that of       
standard generally relativistic theories, where we must allow in the          
action, in addition to the                                                    
ordinary measure of integration $\sqrt{-g}d^{4}x$, another one,               
$\Phi d^{4}x$, where $\Phi$ is a density built out of degrees of freedom      
independent of the metric.                                                    
                                                                              
        For example, given 4-scalars $\varphi_{a}$ (a =                      
1,2,3,4), one can construct the density                                       
\begin{equation}                                                              
\Phi =  \varepsilon^{\mu\nu\alpha\beta}  \varepsilon_{abcd}                   
\partial_{\mu} \varphi_{a} \partial_{\nu} \varphi_{b} \partial_{\alpha}       
\varphi_{c} \partial_{\beta} \varphi_{d}                                      
\end{equation}                                                                
                                                                              
        One can allow both geometrical                                        
objects to enter the theory and consider \cite{modmes}                                  
\begin{equation}                                                              
S = \int L_{1} \Phi  d^{4} x  +  \int L_{2} \sqrt{-g}d^{4}x                   
\end{equation}                                                                
                                                                              
         Here $L_{1}$ and $L_{2}$ are                                         
$\varphi_{a}$  independent. There is a good reason not to consider            
mixing of  $\Phi$ and                                                         
$\sqrt{-g}$ , like                                                            
for example using                                                             
$\frac{\Phi^{2}}{\sqrt{-g}}$. This is because (2) is invariant (up to the inte
divergence) under the infinite dimensional symmetry                           
$\varphi_{a} \rightarrow \varphi_{a}  +  f_{a} (L_{1})$                       
where $f_{a} (L_{1})$ is an arbitrary function of $L_{1}$ if $L_{1}$ and      
$L_{2}$ are $\varphi_{a}$                                                     
independent. Such symmetry (up to the integral of a total divergence) is      
absent if mixed terms are present.                                            
                                                                              
        We will study now the dynamics of a scalar field $\phi$ interacting   
with gravity as given by the action (2) with \cite{decl,cosco,gold}                              
\begin{equation}                                                              
L_{1} = \frac{-1}{\kappa} R(\Gamma, g) + \frac{1}{2} g^{\mu\nu}               
\partial_{\mu} \phi \partial_{\nu} \phi - V(\phi),  L_{2} = U(\phi)           
\end{equation}                                                                
                                                                              
\begin{equation}                                                              
R(\Gamma,g) =  g^{\mu\nu}  R_{\mu\nu} (\Gamma) , R_{\mu\nu}                   
(\Gamma) = R^{\lambda}_{\mu\nu\lambda}, R^{\lambda}_{\mu\nu\sigma} (\Gamma) = 
\Gamma^{\lambda}_{\mu\nu,\sigma} - \Gamma^{\lambda}_{\mu\sigma,\nu} +                          
\Gamma^{\lambda}_{\alpha\sigma}  \Gamma^{\alpha}_{\mu\nu} -                   
\Gamma^{\lambda}_{\alpha\nu} \Gamma^{\alpha}_{\mu\sigma}.                     
\end{equation}                                                                
                                                                              
        In the variational principle $\Gamma^{\lambda}_{\mu\nu},              
g_{\mu\nu}$, the measure fields scalars                                       
$\varphi_{a}$ and the  scalar field $\phi$ are all to be treated              
as independent variables.                                                     
        If we perform the global scale transformation ($\theta$ =             
constant)                                                                     
\begin{equation}                                                              
g_{\mu\nu}  \rightarrow   e^{\theta}  g_{\mu\nu}                              
\end{equation}                                                                
then (2), with the definitions (3), (4), is invariant provided  $V(\phi)$     
and $U(\phi)$ are of the                                                      
form                                                                          
\begin{equation}                                                              
V(\phi) = f_{1}  e^{\alpha\phi},  U(\phi) =  f_{2}                            
e^{2\alpha\phi}                                                               
\end{equation}                                                                
and $\varphi_{a}$ is transformed according to                                 
$\varphi_{a}   \rightarrow   \lambda_{a} \varphi_{a}$                         
(no sum on a) which means 
$\Phi \rightarrow \biggl(\prod_{a} {\lambda}_{a}\biggr) \Phi \equiv \lambda   
\Phi $                                                                        
such that                                                                     
$\lambda = e^{\theta}$                                                        
and                                                                           
$\phi \rightarrow \phi - \frac{\theta}{\alpha}$. In this case we call the     
scalar field $\phi$ needed to implement scale invariance "dilaton".

\subsection{Equations of Motion}
        Let us consider the equations which are obtained from                 
the variation of the $\varphi_{a}$                                            
fields. We obtain then  $A^{\mu}_{a} \partial_{\mu} L_{1} = 0$                
where  $A^{\mu}_{a} = \varepsilon^{\mu\nu\alpha\beta}                         
\varepsilon_{abcd} \partial_{\nu} \varphi_{b} \partial_{\alpha}               
\varphi_{c} \partial_{\beta} \varphi_{d}$. Since                              
det $(A^{\mu}_{a}) =\frac{4^{-4}}{4!} \Phi^{3} \neq 0$ if $\Phi\neq 0$.       
Therefore if $\Phi\neq 0$ we obtain that $\partial_{\mu} L_{1} = 0$,          
 or that                                                                      
$L_{1}  = M$,                                                                 
where M is constant. This constant M appears in a self-consistency            
condition of the equations of motion                                          
that allows us to solve for $ \chi \equiv \frac{\Phi}{\sqrt{-g}}$             
                                                                              
\begin{equation}                                                                 
\chi = \frac{2U(\phi)}{M+V(\phi)}.                                            
\end{equation}                                                                
                                                                              
        To get the physical content of the theory, it is convenient to go     
to the Einstein conformal frame where                                         
\begin{equation}                                                              
\overline{g}_{\mu\nu} = \chi g_{\mu\nu}                                       
\end{equation}                                                                
and $\chi$  given by (7). In terms of $\overline{g}_{\mu\nu}$   the non       
Riemannian contribution (defined   as                                         
$\Sigma^{\lambda}_{\mu\nu} =                                                  
\Gamma^{\lambda}_{\mu\nu} -\{^{\lambda}_{\mu\nu}\}$                           
where $\{^{\lambda}_{\mu\nu}\}$   is the Christoffel symbol),                 
disappears from the equations, which can be written then in the Einstein      
form ($R_{\mu\nu} (\overline{g}_{\alpha\beta})$ =  usual Ricci tensor)        
\begin{equation}                                                              
R_{\mu\nu} (\overline{g}_{\alpha\beta}) - \frac{1}{2}                         
\overline{g}_{\mu\nu}                                                         
R(\overline{g}_{\alpha\beta}) = \frac{\kappa}{2} T^{eff}_{\mu\nu}             
(\phi)                                                                        
\end{equation}                                                                
where                                                                         
\begin{equation}                                                              
T^{eff}_{\mu\nu} (\phi) = \phi_{,\mu} \phi_{,\nu} - \frac{1}{2} \overline     
{g}_{\mu\nu} \phi_{,\alpha} \phi_{,\beta} \overline{g}^{\alpha\beta}          
+ \overline{g}_{\mu\nu} V_{eff} (\phi),                                       
V_{eff} (\phi) = \frac{1}{4U(\phi)}  (V+M)^{2}.                               
\end{equation}                                                                
        If $V(\phi) = f_{1} e^{\alpha\phi}$  and  $U(\phi) = f_{2}
e^{2\alpha\phi}$ as                                                           
required by scale invariance, we obtain from (10)                             
\begin{equation}                                                              
        V_{eff}  = \frac{1}{4f_{2}}  (f_{1}  +  M e^{-\alpha\phi})^{2}        
\end{equation}                                                                
                                                                              
        Since we can always perform the transformation $\phi \rightarrow      
- \phi$ we can                                                                
choose by convention $\alpha > O$. We then see that as $\phi \rightarrow      
\infty, V_{eff} \rightarrow \frac{f_{1}^{2}}{4f_{2}} =$ const.                
providing an infinite flat region. Also a minimum is achieved at zero         
cosmological constant , without fine tuning 
for the case $\frac{f_{1}}{M} < O$ at the point         
$\phi_{min}  =  \frac{-1}{\alpha} ln \mid\frac{f_1}{M}\mid $. Finally,        
the second derivative of the potential  $V_{eff}$  at the minimum is          
$V^{\prime\prime}_{eff} = \frac{\alpha^2}{2f_2} \mid{f_1}\mid^{2} > O$        

\section{Interpretations, Generalizations and Physical Applications of 
the Model}
There are many interesting issues that one can raise here. The first one      
is of course the fact that a realistic scalar field potential, with           
massive excitations when considering the true vacuum state, is achieved in    
a way consistent with the idea of scale invariance. An interesting point
to be made concerning this is that even though spontaneous symmetry breaking
has taken place, no Goldstone boson appears nevertheless 
 \cite{gold} when analyzing the theory in its ground state. This 
interesting and unusual effect is due to the fact that although a locally
conserved current can be defined (from Noether's theorem), this still 
does not lead to a globally conserved charge, because the currents have an
infrared singular behavior that causes scale charge to leak to 
infinity \cite{gold}.

        The second point to be raised is that since there is an infinite      
region of flat potential for $\phi \rightarrow \infty$, we expect a slow      
rolling                                                                    
inflationary scenario to be                                               
viable, provided the universe is started at a sufficiently large value of     
the scalar field $\phi$ for example.                                                      
It is also very interesting to notice that the theory can be related to the
induced gravity theory of Zee \cite{Zee}, defined by the action,                              
                                                                                                              
\begin{equation}                                                              
S  = \int \sqrt{-g} (-  \frac {1}{2} \epsilon \varphi^{2} R +
\frac{1}{2} g^{\mu\nu} \partial_{\mu} \varphi \partial_{\nu} \varphi
-\frac {\lambda}{8} ( \varphi^{2} - \eta^{2} )^{2} ) d^{4} x                    
\end{equation}                                                                
                                                                              
Here it is assumed that the second order formalism is used, i.e.              
$R=R(g)=$ usual Riemannian scalar curvature defined in terms of $ g_{\mu\nu}$.
Notice that if  $ \eta = 0 $, the action is        
invariant under the global scale transformation                               
$g_{\mu\nu}  \rightarrow   e^{\theta}  g_{\mu\nu}$, $\varphi \rightarrow
e^{-\frac{\theta}{2}} \varphi$, but a finite induced Newton's constant
is defined only if $ \eta $ is non vanishing. Then defining ($2k^{2} =\kappa$) 
$\overline{g}_{\mu\nu} = 
k^{2} \epsilon \varphi^{2} g_{\mu\nu}$ and the scalar field                                                                                                 
$ \phi = \frac {1}{k} \sqrt{ 6 + \frac {1}{\epsilon} } ln \varphi $,                            
one can then show that the induced gravity  model is equivalent to standard    
General  Relativity (expressed in terms of $\overline{g}_{\mu\nu}$ )
 minimally coupled to the scalar field $\phi$ which has a potential \cite{K-i} , 
$V_{eff}  = \frac{\lambda}{8 k^{4} \epsilon^{2}}  (1  - 
 \eta^{2} e^{-2\sqrt{ \frac {\epsilon}{1 + 6 \epsilon} } k \phi})^{2}$                                                            
 which is exactly the form (11) with 
$\alpha = 2\sqrt{ \frac {\epsilon}{1 + 6 \epsilon}} k $ (in Ref.6, $k$ is called
$\kappa$). 
The induced gravity model (12) is quite 
successful from the point of view of its applications to inflation and it 
has been studied by a number of authors in this context \cite{infla}. 
Notice that the induced gravity model is not 
consistent with scale invariance for a non vanishing $\eta$, while the theory 
developed here, which leads to the induced gravity model after ssb, 
has been constructed
starting with scale invariance as a fundamental principle.                                              
 
           Furthermore, one can consider this model as suitable for the      
present day universe rather than for the early universe, after we suitably    
reinterpret the meaning of the scalar field  $\phi$. This can provide a long  
lived almost constant vacuum energy for a                                     
long period of time, which can be small if $f_{1}^{2}/4f_{2}$ is              
small. Such small energy                                                      
density will eventually disappear when the universe achieves its true         
vacuum state.                                                                 
                                                                              
        Notice that a small value of $\frac{f_{1}^{2}}{f_{2}}$   can be       
achieved if we let $f_{2} >> f_{1}$. In this case                             
$\frac{f_{1}^{2}}{f_{2}} << f_{1}$, i.e. a very small scale for the           
energy                                                                        
density of the universe is obtained by the existence of a very high scale     
(that of $f_{2}$) the same way as a small fermion mass is obtained in the     
see-saw mechanism \cite{seesaw} from the existence also of a large mass scale. It     
can be shown also that if we take $f_{2} >> f_{1}$, so that the vacuum 
energy is small, this also produces a drastic suppression in the fermion 
particle masses (everything can be done consistently with scale invariance,
see Ref.3) and a possible correlation between the lightest particle mass 
and the vacuum energy of the universe is argued for (see the transparencies 
of this talk which can be found in the web site of this conference). The scale 
invariant way of introducing fermion masses has also implications concerning
the "cosmic coincidences" problem (see Ref.3).
                                  
Finally, this kind of theories can naturally provide a dynamics that 
interpolates between a high energy density (associated with inflation) 
and a very low energy density (associated with the present universe). 
For this consider two scalar fields $\phi_{1}$ and $\phi_{2}$, with normal kinetic terms coupled to the measure
 $\Phi$ as it has been done with the simpler model of just one scalar field.
Introducing for $\phi_{1}$ a potential $V_{1}(\phi_{1}) = 
a_{1} e^{\alpha_{1}\phi_{1}}$ that couples to $\Phi$  and another 
$U_{1}(\phi_{1}) = b_{1}e^{2\alpha_{1} \phi_{1}}$  
 that couples to $\sqrt{-g}$ as required by scale invariance and the potential
for $\phi_{2}$, 
 $V_{2}(\phi_{2}) = a_{2} e^{\alpha_{2}\phi_{2}}$  that couples to $\Phi$  and another
 $U_{2}(\phi_{2}) = b_{2}e^{2\alpha_{2}\phi_{2}}$  
that couples to $\sqrt{-g}$, we arrive (after going through the same steps
as those explained in the model with just one scalar, i.e. solving the constraint and
going to the Einstein frame) at 
the the effective potential
 
\begin{equation}                                                              
        V_{eff}  = 
\frac{(V_{1}(\phi_{1})  + V_{2}(\phi_{2}) + M )^{2}}{4(U_{1}(\phi_{1})  + U_{2}(\phi_{2}))} 
\end{equation}                                                                

which introduces interactions between $\phi_{1}$  and $\phi_{2}$, although no
interactions appeared in the original action (i.e. no direct couplings 
appeared). If we take then $\alpha_{1} \phi_{1}$ very big while $\phi_{2}$
is fixed, then $ V_{eff} $ approaches the constant value 
$\frac{a_{1}^{2}}{4b_{1}}$ while if we take $\alpha_{2} \phi_{2}$ 
to be very big while  $\phi_{1}$  is kept fixed, then $ V_{eff} $ 
approaches the constant value 
$\frac{a_{2}^{2}}{4b_{2}}$. One of these flat regions of the potential can be
associated with a very high energy density, associated with inflation and the
other can be very small and associated with the energy density of the present 
universe. The effective potential (13) provides therefore a dynamics that 
interpolates naturally between the inflationary phase and the 
present slowly accelerated universe.

\end{document}